# AlN Nanowire-Based Vertically Integrated Piezoelectric Nanogenerators

N. Buatip[1], T. Auzelle[2], P. John[2], S. Rauwerdink[2], M. Sohdi[1], M. Salaün[1], B. Fernandez[1], E. Monroy[3], D. Mornex[1], C. R. Bowen[4], R. Songmuang[1*]

[1] Université Grenoble Alpes, CNRS, Grenoble INP, Institut Néel, 38000 Grenoble, France
[2] Paul-Drude-Institut für Festköperelektronik, Leibniz-Institut im Forschungsverbund Berlin e.V. Hausvogteiplatz 5-7, 10117 Berlin, Germany
[3] Université Grenoble-Alpes, CEA, Grenoble INP, IRIG, PHELIQS, F-38000, Grenoble, France
[4] Department of Mechanical Engineering, University of Bath, BA2 7AY Bath, UK
*Corresponding author: rudeesun.songmuang@neel.cnrs.fr

Abstract

In this study, detailed analysis of the direct piezo-response of AlN nanowire-based vertically integrated nanogenerators (VINGs) is undertaken as a function of mechanical excitation frequency. We show that the piezo-charge, piezo-voltage, and impedance measured at the same position of the devices can be directly correlated through an equivalent circuit model, in the whole frequency range of investigation. Our presented results are utilized to determine the performance figures of merit (FoM) of nanowire-based VINGs, namely the piezoelectric voltage constant ($g$) for sensing, and the product $d \cdot g$ for energy harvesting, where $d$ is the piezoelectric charge constant. By comparison of these metrics with those of freestanding single crystal GaN and quartz substrates, as well as sputtered AlN thin films, we suggest that the nanowires can outperform their rigid counterparts in terms of mechanical sensing and energy generation. This work provides experimental guidelines for understanding the direct piezo-characteristics of VINGs and facilitates a quantitative comparison between nanostructured piezoelectric devices fabricated using different materials or architectures.

1. Introduction

Piezoelectric nanogenerators, which are capable of converting ambient mechanical energy into electrical energy, have emerged as a research hotspot in recent decades [1-6]. These devices offer promising prospects for driving low-power electronics and serving as self-powered sensors in forthcoming self-sustained and smart systems. The relevance of such devices extends to various domains including the Internet of Things, human-machine interaction, and wearable technology. The pursuit of high-performance piezoelectric nanogenerators emphasizes the necessity to explore piezoelectric materials that demonstrate flexibility, biocompatibility, and high energy conversion efficiency. It was suggested that large-aspect-ratio piezoelectric nanowires which are homogeneous distributed in polymer matrix, could be the optimal structure that offers substantial energy output [7].

In this regard, vertically-aligned nanowires of wurtzite piezoelectric semiconductors, including ZnO and III-Nitrides, stand out as promising active elements for piezoelectric nanogenerators due to their geometry that offers superior flexibility and strain sensitivity compared to traditional high stiffness, and brittle piezoelectric materials. Although wurtzite nanowires show lower piezoelectric coefficients than conventional piezoelectric materials (for example, Lead Zirconate Titanate (PZT)), their growth approach generally provides a high degree of control and tunability. In addition, wurtzite semiconductors exhibit a permanent polarization along the [0001] crystallographic direction without the need for a poling process, unlike for ferroelectric materials.

Among these wurtzite semiconductors, AlN, with a reasonable piezoelectric coefficient of 5.3 pC/N [8-10], possesses the largest bandgap of 6.2 eV, resulting in a highly insulating feature which minimizes screening effects that occur in unintentionally n-type doped ZnO [11-13]. Considering the attractive intrinsic material properties of AlN, coupled with the mechanical flexibility of nanowires, AlN nanowires are expected to enhance the nanogenerator performance. However, AlN nanowire-based nanogenerators have not been widely explored in contrast to the extensive research on ZnO and GaN nanowires. This fact likely stems from the limited availability of AlN nanowires, whose synthesis is rather demanding [14]. Recently, it was shown that high crystal quality and vertically-aligned AlN nanowires can be achieved using plasma-assisted molecular beam epitaxy (PAMBE) [15-16]. In this work, we implement such AlN nanowires in vertically integrated nanogenerator (VING) architecture [17] and perform detailed investigations on their direct piezoresponse characteristics, as well as evaluate their potential for mechanical sensing and energy generating applications.

In spite of a growing body of literature that reports an increase in signal output from nanogenerators using a variety of nanomaterials and architectures [3-6], the ability to accurately evaluate and compare these devices remains a topic of ongoing discussion [18-23]. This is attributed to the often-limited details of in-house measurement setups used in the literature and the absence of a standardized characterization protocol. The detected signals of nanogenerators are typically in the form of voltage and current spikes, which are produced during rapid impulsive pressing or bending of the piezoelectric devices [1-2, 24-26]. While such measurements can offer useful information related to device efficiency, they usually do not provide a comprehensive and comparable description of performance and operational characteristics across different devices. Furthermore, only few works have paid attention on measuring the generated piezoelectric charge [13, 27-28], unlike the voltage signal, even though the charge is the primary output and a crucial evaluation metric [29]. Therefore, to assess the potential applications of nanogenerators and their integration compatibility with other electronic components, more detailed analyses of the piezoelectric response are required.

In this work, our characterisation setup is designed based on the Berlincourt principle [30], which is used in commercial systems for measuring the piezoelectric charge coefficients of materials [31-33]. The Berlincourt concept involves applying a controlled force to the device under test (DUT) and detecting its mechanically induced electrical signal. The piezoresponse (i.e., charge and voltage) is investigated as a function of frequency of the mechanical excitation. Electrical measurements, such as current-voltage and impedance analyses, are also carried out to gain an insight into the resistance and capacitance of the devices. The obtained data are used for constructing an equivalent circuit that quantitatively describes the direct piezoelectric response behaviours as a function of excitation frequency. The advantage of our setup is that it measures the piezo-charge, piezo-voltage, and device impedance at the same position of the device, thereby greatly facilitating signal interpretation and the correlation of each property. Such studies have rarely been reported to date.

To ensure the reliability and accuracy of the experimental setup, we first analysed the piezo-response of well-known reference piezoelectric materials, namely a freestanding GaN substrate and an x-cut quartz substrate. We also investigated a sputtered AlN thin film on Si substrate for comparison. Subsequently, we applied our measurement protocol to evaluate vertically-aligned GaN and AlN nanowires which are encapsulated in polydimethylsiloxane (PDMS) and fabricated into a VING architecture. We demonstrate that a classical equivalent circuit model for piezo-sensing devices can effectively capture the characteristics of the piezoelectric response of the nanowire-based VINGs. From our measurements, we can obtain the piezoelectric charge constant ($d$), allowing us to extract the piezoelectric voltage constant ($g$) from $d/\epsilon$ and the $d \cdot g$ product which corresponds to $d^2/\epsilon$. These metrics represent the figures of merit (FoM) for piezoelectric sensors and energy harvesters, respectively [34-35]. Our observations reveal that the $g$ coefficient of the AlN nanowire-based VINGs can exceed that of the conventional reference samples, indicating the superior suitability and advantages of the nanowires for mechanical sensing applications. When considering the $d \cdot g$ value which represents the volume density of energy generated by an applied stress, the result also suggests that the AlN nanowire-based VINGs can provide the output power higher than that of the reference materials, indicating untapped potential for energy generating devices.

## 2. Experimental detail

*2.1 Nanowire growth and device fabrication*

To investigate single crystal piezoelectric materials, a freestanding (0001) Fe-doped GaN semi-insulating substrate (6 × 10 mm$^2$) with a thickness of 475 ± 0.25 µm was purchased from Kyma Technology. The resistivity of the substrate was >10$^6$ Ω·cm. For an x-cut quartz, a substrate (10 × 10 mm$^2$) with a thickness of 500 µm was purchased from MTI corporation. An additional planar sample consists of a 300 nm of AlN layer grown on a highly doped Si(111) substrate using radio-frequency-magnetron sputtering at 300 °C. In all these cases, a 10-nm Ti/100-nm Au bilayer was deposited on the sample surface to serve as an electrical contact.

For the study of the nanowire-based nanogenerators, vertically aligned Al-polar AlN and N-polar GaN nanowires were grown using plasma-assisted molecular beam epitaxy (PAMBE). The Al-polar AlN nanowires were grown on a 400 nm TiN buffer layer on a (0001) sapphire substrate. The TiN is highly conductive and can be used as a bottom electrode for nanowire devices. Exact details of their growth and structural analyses were described elsewhere [15-16, 36]. The N-polar GaN nanowires were grown on a highly doped n-type Si(111) substrate, with a thin AlN buffer layer. In this case, the conductive Si substrate was used as the bottom contact of the nanowires. Both types of nanowires have diameters and lengths in the range of 50-100 nm and 500-1000 nm, respectively, with nanowire densities ranging from $10^9$ to $10^{10}$ cm$^{-2}$, as determined using scanning electron microscopy (SEM). Figures 1(a)-(b) show typical top and side view SEM images of the as-grown AlN nanowires. The polarity of the nanowires was verified by piezo-response force microscopy [37] and transmission electron microscopy on the samples grown under identical conditions [15,38].

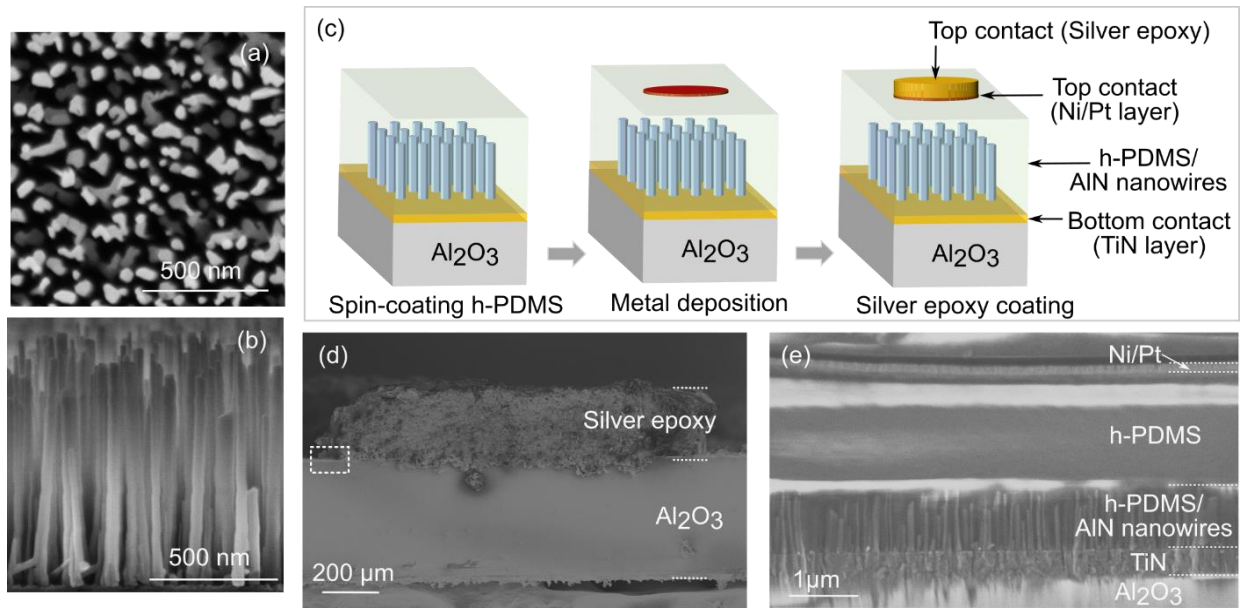

*Figure 1: (a) Top and (b) side view SEM images of the as-grown AlN nanowires. (c) Schematic illustration showing the fabrication process of nanowire-based VINGs (see the section 2.1 for detail description). The schematic is not drawn to scale. (d) Side view SEM image of AlN nanowire-based VING and (e) magnified view of the square area in (d).*

To investigate and characterise the direct piezo-response of the nanowire ensemble, we fabricated a conventional nanowire VING structure, as shown in the schematic in Figure 1(c). The surfaces of GaN and AlN nanowires were cleaned using oxygen plasma via a PLASSYS reactive ion etching system (10 W for 15 seconds). Subsequently, the nanowires were encapsulated in a hard-polydimethylsiloxane (h-PDMS) [39] using a spin coating technique. The samples were then baked at 80 °C for 30 minutes under a vacuum of 90 mbar. Before electrode deposition, the encapsulated layer was activated using oxygen plasma (10 W for 15 seconds) to enhance the bonding between the h-PDMS and metal contact [40]. Then, a 5-nm Ni/100-nm Pt layer was evaporated onto the sample to serve as a top electrode. Subsequently a silver epoxy (M.G. Chemicals, 8330D) with a thickness of a few hundred micrometres was covered over the top electrode to improve device stability. Figures 1(d)-(e) show large-scale and zoomed-in SEM images of a typical AlN nanowire-based VING fabricated in this work. The latter image reveals a 2-µm thick layer of h-PDMS remaining on top of the encapsulated nanowires.

*2.2 Direct piezo-response measurements and electrical characterisation*

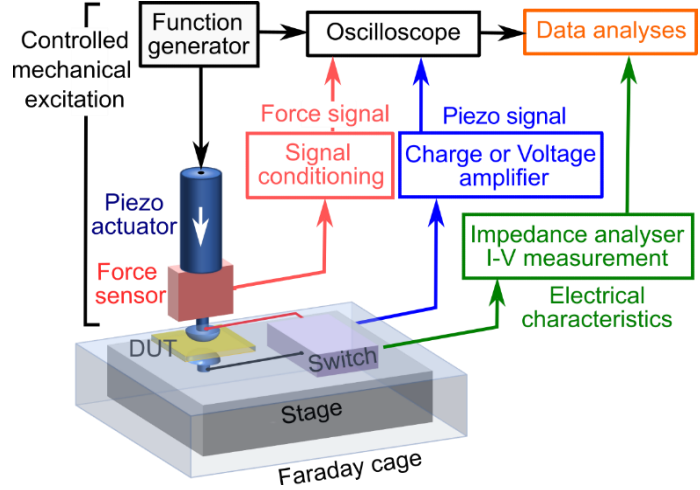

Figure 2: Schematic of the measurement setup for piezo-voltage, piezo-charge, and electrical characteristics of the piezoelectric devices. (see the section 2.2 for detail description).

A schematic of our measurement setup is shown in Figure 2, where a dynamic force with a controlled frequency and waveform was applied to the DUTs through a piezoelectric actuator (Physik Instrumente, PI840.30). The applied force was measured using the force sensor (Futek LSB201). In this work, we applied a sinusoidal force for excitation, which is represented by $F(f_{exc}, t) = F_{dc} + F_{ac} \cdot \sin(2\pi f_{exc} \cdot t + \phi_F)$, where $f$ and $\phi_F$ are the frequency and phase of the excitation, respectively. The amplitude of the applied force ($F_{ac}$) ranges from 300 to 500 mN, and the excitation frequency ($f_{exc}$) was swept from 0.3 to 150 Hz. A preload ($F_{dc}$) with a value between 2 and 4 N was applied to ensure electrical and mechanical stability. This preload does not affect the dynamic piezoelectric response characteristics of the rigid samples.

The electronics for piezo-response signal acquisition consists of a charge amplifier (Kistler 1505A) with an operating frequency that ranged from 0.16 Hz to 30 kHz and a low-frequency voltage amplifier with an input resistance ($R_{in}$) of 1 TΩ and an input capacitance ($C_{in}$) of 18 pF (FEMTO, DLPVA-100-F), which can be operated from direct current (DC) to 200 kHz. After measuring the piezoelectric signal, an impedance measurement was performed at the same position under the same static preload using an LCR meter (Keysight 4980AL) with an excitation voltage of 2 V and frequency ranging from 20 Hz to 1 MHz. The capacitance ($C_{pz}$) and resistance ($R_{pz}$) values of the DUTs were measured using a parallel equivalent circuit model. The current-voltage (*I-* $V$) characteristics of the DUTs were measured using a source meter unit (Keithley 2636B). The DC resistance ($R_{dc}$) was derived from the I-V curve obtained by applying the bias ranging from -1 V to +1 V.

## 3. Results and discussion

*3.1 Piezoelectric response of a reference GaN bulk substrate*

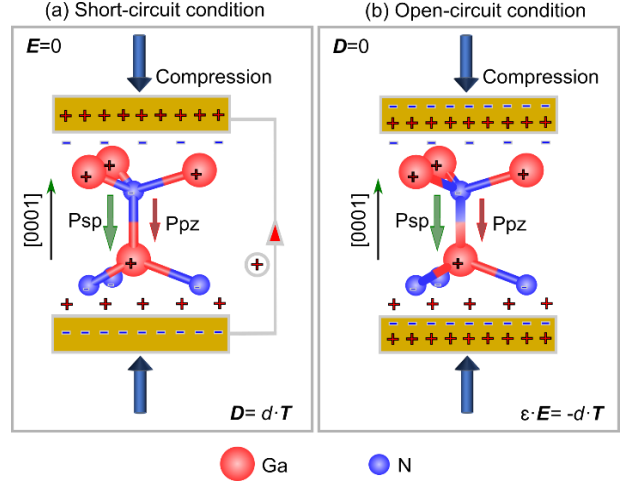

Figure 3 Schematics illustrating a Ga-polar GaN crystal subjected to a compressive force along its c-axis under (a) short-circuit and (b) open-circuit conditions.

Figures 3(a)-(b) illustrate the crystal structure of wurtzite GaN and its polarization. As a polar crystal, GaN exhibits a spontaneous polarization ($P_{sp}$), an inherent polarization that occurs at zero external stress. The sign of $P_{sp}$ of Ga-polar GaN is *negative* in the [0001] direction. When an external stress ($T$) is applied, the centres of symmetry of the positive and negative ions shift further away from each other, thereby generating a piezoelectric polarization $P_{pz} = d \cdot T$. The applications of a compressive stress along the *c*-axis of the Ga-polar GaN crystal induces a *negative* $P_{pz}$ [8]. Following the constitutive equation for a direct piezoelectric effect [41,42], the electric displacement ($D$) is described by equation (1) as follows.

$$D = \epsilon \cdot E + d \cdot T \qquad (1)$$

where $E$ is the electric field within the material and $\epsilon$ is the material dielectric permittivity. Under a short-circuit condition ($E = 0$), equation (1) is reduced to $D = d \cdot T$. In this scenario, $P_{pz}$ is compensated by the electric field from the true charges on the electrode, resulting in the vanishing of the resultant field in the material; see Figure 3(a). In contrast, no true charge appears on the electrodes under open-circuit conditions ($D = 0$), as shown in Figure 3(b). Hence, the electric field from the surface charges in the piezoelectric crystal follows the relation: $\epsilon \cdot E = -d \cdot T$.

The value and sign of $P_{pz}$ can be deduced from the charges flowing through the external circuit, e.g. using voltage or charge amplifiers [30,32,33,43,44]. In the charge detection circuit, an operational amplifier (op-amp) compensates for the generated piezoelectric charges in the DUTs by charging the feedback capacitance to maintain a zero-voltage difference between the two input terminals, thus applying the short circuit condition (see also Figure S1 in the supporting information). In our measurement configuration, the force is exerted in the z-direction along the c-axis of the GaN substrate. Considering only the dynamic response, the electrical displacement ($D_z$) in the time domain can be quantified by equation (2).

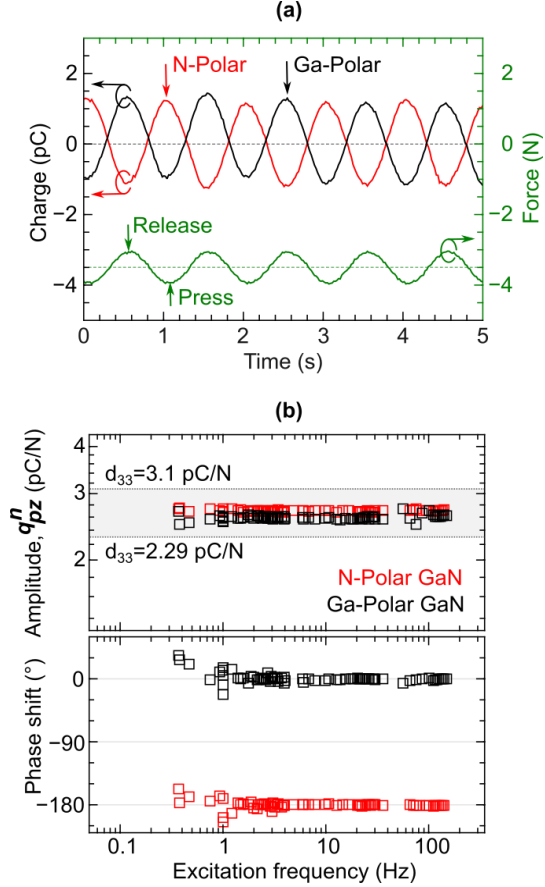

Figure 4: (a) Charge signals from the Ga-and N-polar surfaces of a GaN semi-insulating substrate during a sinusoidal force excitation at 1 Hz, shown in black and red solid lines, respectively. (b) Corresponding normalized charge amplitude ($q_{pz}^n$) and phase response as a function of excitation frequency from 0.3 to 150 Hz. Grey shade represents the range of $d_{33}$ of GaN from the literature, i.e. $d_{33} = 3.1$ pC/N [10,44] and $d_{33} = 2.29$ pC/N [8].

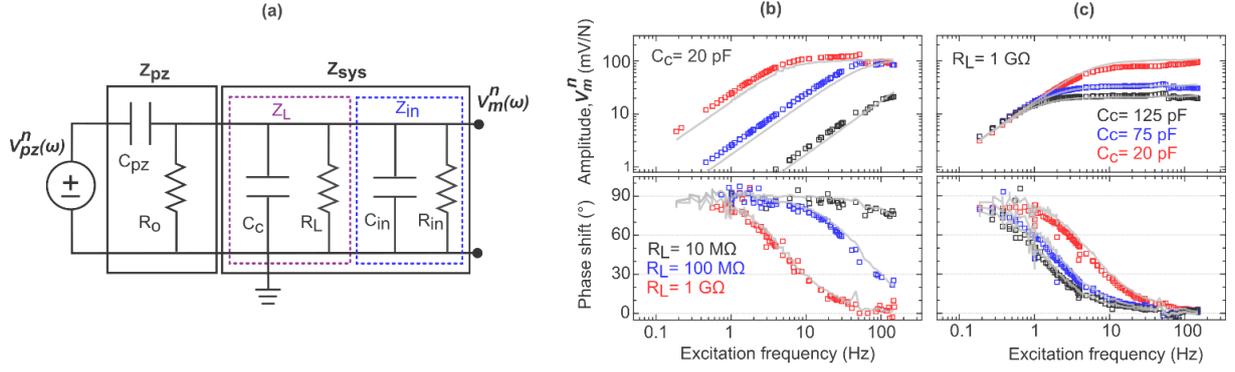

*Figure 5: (a) Equivalent circuit of a piezoelectric material connected to the voltage detection circuit. $C_{in}$ and $R_{in}$ of the voltage amplifier are 18 pF and 1 T$\Omega$, respectively, while $C_{pz}$ and $R_o$ of the studied DUTs are presented in Table 1. (b)-(c) Normalized voltage amplitude and phase response from Ga-polar GaN semi-insulating substrate measured using various $R_L$ and $C_c$. Grey solid lines in (b) and (c) represent the amplitude and the phase response calculated from the equivalent circuit in (a), using $C_{pz} = 8$ pF and $R_o >$10 G$\Omega$.*

$$D_z(t) = \frac{d_{33,eff}}{A_F} \cdot F(t) = \frac{Q_z(t)}{A_Q} \quad (2)$$

where $A_F$ is the area where the force is applied, and $A_Q$ is the area where the charges are developed. These areas are generally of the same size. The parameter $Q_z(t)$ is the generated piezoelectric charge on the <0001> surface of the GaN and $F(t)$ is the applied force, while $d_{33,eff}$ is the effective piezoelectric charge coefficient along the stressed direction. As the excitation frequency is sufficiently low, we can impose a quasi-static condition. The $d_{33,eff}$ is determined by dividing the charge amplitude by the force amplitude [32,33].

Figure 4(a) presents the charge signals from the Ga-and N-polar surfaces of a GaN semi-insulating bulk substrate, measured using the charge amplifier. The sinusoidal force excitation, represented by the green solid line, has an amplitude ($F_{ac}$) of 440 mN and a frequency ($f$) of 1 Hz. The minimum point of the force signal represents the compressive force on the sample surface generated by the piezo-actuator, whereas the maximum force related to the situation when the piezo-actuator was released from the substrate. The negative DC offset in the force signal results from the application of a preload of 3.5 N that presses on the sample surface during the measurement.

The black solid line curve in Figure 4(a) reveals that negative charges were generated on the Ga-polar surface of the GaN substrate when a compressive force was applied. Conversely, positive charges appear when the force was released from the surface. When the sample was physically inverted, positive charges were detected under the compressive force on the N-polar surface. This observation is consistent with the sign of $P_{pz}$ of GaN, thereby evidencing the piezoelectric nature of the signals. Substrate bending which could strongly influence the piezoelectric signal [37,45], is ruled out since the force is applied symmetrically on both sides of the samples.

Figure 4(b) illustrates the charge signal analysed in relation to the mechanical excitation and plotted as a function of frequency, ranging from 0.3 to 150 Hz. The upper panel of this figure shows the charge amplitude normalized by the force amplitude ($q_{pz}^n$), corresponding to the $d_{33,eff}$ of the investigated GaN substrate. The obtained $d_{33,eff}$ of 2.7 pC/N remains constant within our excitation frequency range and falls within the range of values reported in the literature [8,10,44,46]. Note that this $d_{33,eff}$ is an underestimated value as it includes the clamping effect inside the crystal [37]. The phase shift is calculated by $\phi_Q - \phi_F$, where $\phi_Q$ is the phase of the charge signal, and $\phi_F$ is that of the mechanical excitation. The lower panel of Figure 4(b) shows the 0° and 180° phase-shift for the Ga-polar and N-polar surfaces of GaN substrates, respectively, confirming our interpretation of Figure 4(a). Our measurements on an x-cut quartz substrate (Figure S2 in the supporting information) also provide the $d_{11,\ eff}$ of 2.23 pC/N, which is again consistent with the value reported in the literature [47,48]. The results are summarized in Table 1.

For the piezoelectric voltage measurement, the signal was detected through a load resistor, $R_L$, which was placed between the two input terminals of a voltage amplifier. The piezoelectric material is modelled as a voltage generator connected with a capacitance ($C_{pz}$) and a leakage resistance ($R_o$) [49]. The interfacing cable is represented by the capacitance $C_c$. During a measurement, the finite input impedance ($Z_{in}$) of the op-amplifier must be considered in the equivalent circuit, as presented in Figure 5(a) [23,50]. The variation of the impedance of the voltage detection circuit can significantly affect the detected voltage signal at different frequencies, unlike the case for charge detection [29]. For this reason, the absolute values of piezo-voltages *cannot* be used as an indicator of device performance without detailed description of testing configurations.

The detected voltage amplitude normalized by the force amplitude ($V_m^n(\omega)$) as a function of frequency ($\omega = 2\pi f$) is related to the piezo-voltage generated by the DUTs ($V_{pz}^n(\omega)$) through the transfer function $H(\omega)$ as follows:

$$V_m^n(\omega) = H(\omega) \cdot V_{pz}^n(\omega) \qquad (3)$$

In the case where $R_o$ is significantly large or highly resistive devices, we obtain $H(\omega) = \frac{Z_{sys}}{Z_{sys}+Z_{pz}}$, where $Z_{sys}$ is the total impedance of the detection circuit, and $Z_{pz}$ is the impedance of the DUTs.

We experimentally demonstrate that the piezoelectric voltage depends on the measurement configuration by varying $R_L$ up to 1 GΩ and $C_c$ up to 125 pF where $C_c$ is the capacitance corresponding to the interfacing coaxial cable. The normalized voltage amplitude ($V_m^n(\omega)$) and phase response as a function of frequency, shown in Figures 5(b)-(c), reveal a high-pass filter characteristic, in which the voltage signals with frequencies below a cut-off frequency ($f_{cut\text{-}off}$) are attenuated, and their phase becomes 90° leading the mechanical excitation signal. Above $f_{cut\text{-}off}$, the amplitude reaches a saturation value. In agreement with the behaviour of the charge signal shown in Figure 4(b), the voltage signal is in-phase with the mechanical excitation for the Ga-polar GaN substrate, while it exhibits a 180° shift for the N-polar substrate (not shown). It should be noted that $f_{cut\text{-}off}$ is determined by the $RC$ time constant of the entire circuit.

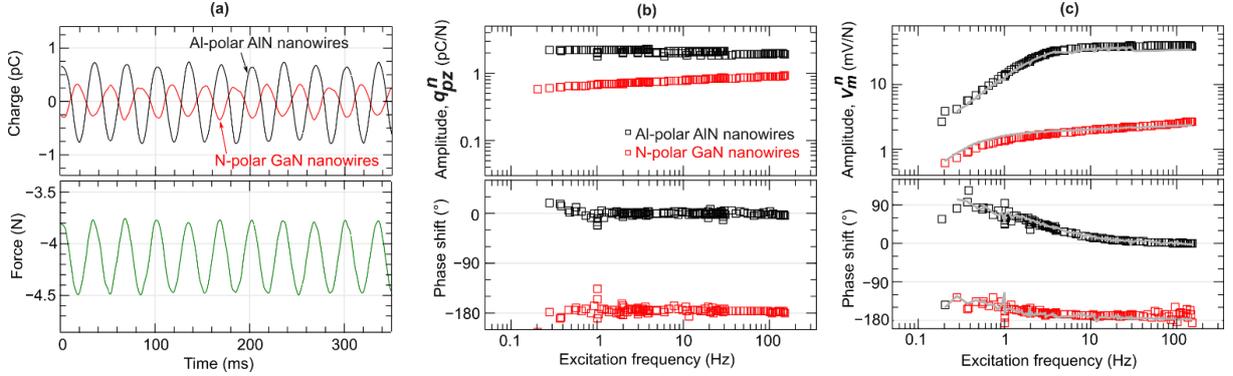

*Figure 6 (a) Charge signal (upper panel) obtained from nanowire-based VINGs, i.e., N-polar GaN nanowires (GaN-NW-01) and Al-polar AlN nanowires (AlN-NW-01), subject to a sinusoidal force excitation of 30 Hz (lower panel). (b) Charge and (c) voltage responses, i.e., normalized amplitude (upper panel) and phase (lower panel), as a function of excitation frequency ranging from 0.3 to 150 Hz. Grey solid lines in (b) and (c) represent the values derived from the equivalent circuit in Figure 5(a), utilizing the device parameters specified in Table 1.*

The normalized piezoelectric charge and voltage signals as a function of frequency can be correlated with the equivalent circuit shown in Figure 5(a) by the following relation:

$$V_{pz}^n(\omega) = \frac{q_{pz}^n(\omega)}{C_{pz}} \approx \frac{2.7}{8} = 0.337 \, V/N \tag{4}$$

We use equation (3) to deduce $V_m^n(\omega)$ from $V_{pz}^n(\omega)$, utilizing the impedances of the detection system and the DUTs. Through impedance measurements of the reference GaN bulk substrate, we obtain a relatively constant capacitance ($C_{pz}$) of 8 pF from 20 Hz up to 1 MHz, with a tangent loss of less than 0.1. The *I-V* characteristic shows a DC resistance ($R_{dc}$) higher than 10 GΩ, which is interpreted as $R_o$ in the equivalent circuit. The parameters of our detection circuit are $C_c$ = 2 pF, $R_L$ = 1 GΩ, $C_{in}$ = 18 pF, and $R_{in}$ = 1 TΩ. The derived $V_m^n(\omega)$ is plotted as grey solid lines in Figures 5(b)-(c), aligning with the amplitude and phase of the piezo-voltage response from all measurement configurations. The normalized saturation voltage ($V_{sat}^n$) obtained above the cut-off frequency can also be estimated by equation (5).

$$V_{sat}^n = \frac{C_{pz}}{C_{pz}+C_c+C_{in}} \cdot V_{pz}^n = \frac{q_{pz}^n}{C_{tot}} \tag{5}$$

where $C_{tot} = C_{pz} + C_c + C_{in}$.

*3.2 Direct piezoresponse analyses of N-polar GaN and Al-polar AlN nanowire-based VINGs*

In this section, we employ the measurement protocol outlined in Section 3.1 to investigate the piezoelectric signals from III-Nitride nanowire-based VINGs. To mitigate the issue of device instability arising from the soft-nature of the PDMS-nanowire composite layer, we reinforced its structural stability and stiffness by covering the top electrode with a thick and more rigid layer of silver epoxy. All presented devices have stable characteristics for over 10,000 cycles across the entire range of excitation frequencies up to 150 Hz. This robust performance ensures the reliability of the analyses.

*Table 1: Summary of the measured $d_{33,\,eff}$ and $V_{sat}$, including electrical parameters ($C_{pz}$, $C_{tot}$, $R_{pz}$, and $R_{dc}$) of the studied DUTs, namely, Ga-GaN semi-insulating substrate, x-cut quartz, sputtered AlN thin films as well as N-polar GaN and Al-polar AlN nanowire-based VINGs. $C_{pz}$ and $R_{pz}$ are the value measured at 20 Hz. $R_{dc}$ is interpreted as $R_o$ in the equivalent circuit shown in Figure 5(a). The extracted values of $C_{tot} \cdot V_{sat}$, $V_{pz}^n$, $V_{sat}^n$, $g_{33,eff}$, and $g_{33,eff} \cdot d_{33,eff}$ are also presented. Theoretical values of $d_{33}$, $g_{33}$ and the product $g_{33} \cdot d_{33}$ of AlN single crystal are also listed for comparison.*

| Device under test (DUTs) | $C_{pz}$ | $C_{tot}$ | $R_{pz}$ | $R_{dc}$ | $d_{33,eff}$ | $C_{tot} \cdot V_{sat}$ | $V_{pz}^n$ | $V_{sat}^n$ | $g_{33,eff}$ | $g_{33,eff} \cdot d_{33,eff}$ |
|---|---|---|---|---|---|---|---|---|---|---|
| | pF | | Ω | | pC/N | | mV/N | | ×10⁻³ Vm/N | ×10⁻¹⁵ m²/N |
| Ga-GaN | 8 | 28 | >10G | >10G | 2.70 | 2.80 | 337 | 100 | 29.3 | 79.1 |
| x-cut quartz* | 10 | 30 | >10G | >10G | 2.23 | 1.95 | 223 | 65 | 55.5 | 123.8 |
| AlN (theoretical)[8] | - | - | - | - | 5.35 | - | - | - | 66.1 | 353.6 |
| Sputtered AlN Layer-01 | 473 | 493 | 1G | >10G | 2.64 | 2.56 | 5.58 | 5.2 | 32.6 | 86.1 |
| GaN-NW-01** | 365 | 385 | 140M | >10G | 0.84 | 0.80 | 2.2 | 2.2 | 31.6 | 26.6 |
| AlN-NW-01 | 37 | 57 | >10G | >10G | 2.34 | 2.22 | 63.2 | 39 | 88.1 | 206.2 |
| AlN-NW-02 | 57 | 77 | >10G | >10G | 2.17 | 2.09 | 38.1 | 27.1 | 81.7 | 177.3 |
| AlN-NW-03 | 54 | 74 | 5.7G | >10G | 2.07 | 1.80 | 38.3 | 24.3 | 77.9 | 161.3 |
| AlN-NW-04 | 26 | 46 | >10G | >10G | 1.27 | 1.20 | 48.8 | 26.5 | 47.8 | 60.7 |
| AlN-NW-05 | 77 | 97 | 1.5G | 8.5G | 0.90 | 0.80 | 11.6 | 7.5 | 33.9 | 30.5 |
| AlN-NW-06** | 255 | 275 | 260M | 1.3G | 0.75 | 0.77 | 2.9 | 3 | 28.2 | 21.2 |
| AlN-NW-07** | 285 | 305 | 15M | 2.2M | 0.65 | 0.60 | 2.3 | 2 | 24.5 | 15.9 |
| ZnO nanowires[27] | - | - | - | - | 3.53 (1Hz) | - | - | - | 64-88 $\epsilon = 4\epsilon_0 - 5.5\epsilon_0$ | 226-311 |
| ZnO nanowires[13] | - | - | - | - | 1.05 (1Hz) | - | - | - | 10.03 ($f_{exc}$=1 Hz) 33.14 ($f_{exc}$=1 kHz) | 10.53 ($f_{exc}$=1 Hz) 34.8 ($f_{exc}$=1 kHz) |
| GaN nanowires[51] (Tapping excitation) | - | - | - | - | - | - | - | 3-100 | - | - |
| GaN nanowires[52] (Step excitation) | - | - | - | - | - | - | - | 13 (V/N) | - | - |

*For x-cut quartz, the $d_{11,eff}$ is measured, and the $\epsilon_{11} = 4.514$ Error! Bookmark not defined.

**Higher capacitance values are attributed to the larger top contact area of these devices, compared to the others.

The upper panel of Figure 6(a) displays the charge response from the VINGs with an active region consisting of N-polar GaN nanowires (GaN-NW-01) and Al-polar AlN nanowires (AlN-NW-01). The sinusoidal excitation force (lower panel) has a preload of 4.25 N, an amplitude of 440 mN and is applied at a frequency of 30 Hz. In Figure 6(b), the normalized charge amplitude ($q_{pz}^n$) and phase in response to the force excitation are plotted as a function of frequency. The $q_{pz}^n$ is interpreted as $d_{33,eff}$ along the c-axis of the wurtzite nanowires, considering that the force was applied in this direction. The extracted values are approximately 2.34 pC/N for AlN-NW-01 and 0.84 pC/N for GaN- NW-01. The observed 0° and 180° phase shifts are consistent with the Al- and N-polarity of the nanowires, perfectly agreeing with the results of the bulk GaN substrate presented in Section 3.1. We further investigate a variety of AlN nanowire-based VINGs and summarize the signal output in Table 1. We systematically found lower values of $d_{33,eff}$ for the VINGs compared to the theoretical $d_{33}$ of bulk AlN, which is attributed to the reduced volume of piezoelectric material in a polymer-nanowire-composite layer[48,53-55]. In addition, the presence of the remaining h-PDMS layer on top of the nanowires leads to a degree of signal loss [55]. Although the studied AlN nanowire-based VINGs were made from the AlN nanowires grown under the same condition, we observed a variation of $d_{33,\ eff}$ in the range of 0.65 - 2.34 pC/N. This discrepancy is ascribed to device geometry fluctuations which depend on various parameters, including contact quality, composite thickness, encapsulation quality, electrical contact quality, and self-assembly nature of the nanowires.

The normalized voltage amplitude $V_m^n(\omega)$ and phase response in Figure 6(c) exhibit a high-pass characteristic, similar to that observed in bulk GaN substrate. The behaviour of $V_m^n(\omega)$, extracted from equation (3) demonstrates good agreement with the detected piezo-voltages for all VINGs studied in this work. As examples, the extracted $V_m^n(\omega)$ from GaN-NW-01 and AlN-NW-01, represented as grey solid lines, are compared with the measured values in Figure 6(c). Table 1 reveals a difference between $V_{sat}^n$ and $V_{pz}^n$ when the device capacitance, $C_{pz}$, is much lower than that of the detection circuit, which is 20 pF in this set of experiments. This result unveils the impact of measurement configuration on piezo-voltage signal. Furthermore, we apply equation (5) using the measured value of $V_{sat}^n$ to calculate $q_{pz}^n$, with the known impedances of our DUTs and detection circuit. These values are consistent with the measured $d_{33,eff}$ in Table 1. This approach provides an alternative method for deducing the $d_{33,eff}$ of the DUTs from their piezo-voltage, without directly measuring the generated charges.

To assess III-Nitride nanowire-based VINGs for sensors and energy harvesters, we determine the corresponding FoM by calculating $g_{33,eff} = \frac{d_{33,eff}}{\epsilon_{33,eff}}$ (for sensing) and the product $g_{33,eff} \cdot d_{33,eff}$ (for energy harvesting), where $\epsilon_{33,eff}$ is the effective dielectric permittivity of the composite layer. In our experiments, the $\epsilon_{33,eff}$ cannot be accurately extracted from the impedance analyses because the device geometry cannot be precisely defined. Hence, we derive the $\epsilon_{33,eff}$ from a simplified model of a composite layer [53, 54]; see the supporting information. The lower $\epsilon_{33,eff}$ of approximately $3 \cdot \epsilon_0$ for the h-PDMS encapsulated AlN and GaN nanowires is able to counterbalance the reduction of $d_{33,eff}$ of the composite layer. Consequently, the $g_{33,eff}$ of the nanowire-based VINGs remain promising compared to the reference bulk materials examined. Table 1 shows that the $g_{33,eff}$ of the AlN nanowire-based VINGs can surpasses those of single crystal GaN and quartz substrates as well as a sputtered AlN layer, indicating their advantage for mechanical sensing applications. Also, the product $d \cdot g$ energy harvesting FoM suggests that the AlN nanowire-based VINGs can offer a higher power output than those of our reference samples. However, it remains to be improved in comparison to the theoretical value of single crystal AlN. It is noteworthy that the derived $\epsilon_{33,eff}$ might be underestimated, leading to an overestimation of $g_{33,eff}$ of the nanowire-based VINGs. In the literature, the $\epsilon_{33,eff} \approx 4\epsilon_0 - 5.5\epsilon_0$ was measured in ZnO nanowire/PMMA composites [27], which have similar material parameters to our DUTs. Despite using the higher value of $\epsilon_{33,eff}$, the deduced $g_{33,eff}$ and $g_{33,eff} \cdot d_{33,eff}$ remains encouraging.

It is important to note that a bending motion of the nanowires is not fully achieved in our experiments, thereby limiting the potential benefits from their inherent flexibility. The bending of nanowires can potentially yield a greater level of charge generation for the same applied force. For example, we have observed the $d_{33,eff}$ of GaN and AlN nanowire-based VINGs up to 3.5 pC/N and 5 pC/N, respectively, which might be ascribed to the bending of the nanowires (see Figure S3 in the supporting information) [56]. These higher-than-the-bulk values of the piezoelectric charge constant should drive the FoMs for both sensing and harvesting beyond those of bulk materials, demonstrating the capability of nanowires as flexible piezoelectric nanogenerators.

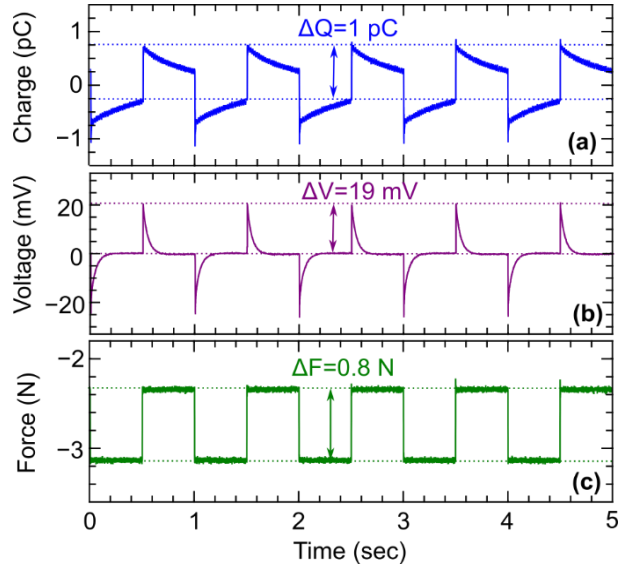

*Figure 7 (a) Charge and (b) voltage signal from the AlN nanowire VINGs (AlN-NW-04) under a step-like force excitation at 1 Hz, as shown in (c).*

Finally, as mentioned in Section 1, most piezo-voltage analyses of nanowire-based VINGs in the literature are performed through intermittent pressing on the devices. Here, in Figures 7(a)-(c), we present the charge and voltage response of the AlN nanowire-based VINGs (AlN-NW-04) plotted as a function of time when subjected to a step-like force excitation. The force presented in Figure 7(c) has a preload of 3.5 N, an amplitude of 400 mN, and a frequency of 1 Hz. Figure 7(a) depicts negative and positive charge signals, when the DUT was compressed or released, respectively. The sign of the signal is consistent with the Al-polarity of the nanowires. Ideally, the charge signal should remain at $-q_{pz}$ or $+q_{pz}$ when the DUT are constantly compressed or released, respectively. The signal decay is caused by a quasi-static detection mode of the charge amplifier, which has the decay time of 1 second in this measurement. By dividing the charge variation ($\Delta Q$) with the force variation ($\Delta F$), we obtain $\frac{\Delta Q}{\Delta F} = $ 1.25 pC/N, which agrees with the $d_{33,\ eff}$ of this device shown in Table 1.

Regarding the piezoelectric voltage signal, voltage spikes ($V_m(t)$) with amplitudes of $-V_0$ or $+V_0$ were observed when the DUT were compressed or released, respectively, as shown in Figure 7(b). The voltage variation ($\Delta V$) and decay time ($\tau$) can be determined from the following equation.

$$V_m(t) = \Delta V_0\, e^{-\frac{t}{\tau}} \tag{6}$$

By fitting equation (6) with the results in Fig.7(b), the normalized voltage amplitude ($\Delta V_0/\Delta F$) and $\tau$ are approximately 24 mV/N and 45 milliseconds, respectively. Since $R_o >> R_L$ in this DUT, a $C_{tot}$ of 45 pF can be calculated from the relation $\tau = R_L \cdot C_{tot}$, allowing us to determine the $d_{33,eff}$ from equation (5). The obtained value is approximately 1.1 pC/N, slightly lower than the value from the charge measurement (1.27 pC/N). This discrepancy is attributed to the finite detection bandwidth of the voltage amplifier and the limited precision of our measurement setup. Nevertheless, it remains a relatively reliable indication of the generated charge from the devices and is an alternative approach for the measurement of piezoelectric charge coefficients. It is noteworthy that the piezoelectric charge response does not significantly change with the step-like force excitation frequency range of 1-30 Hz (see Figure S4 in the supporting information).

In Table 1, we include the $d_{33,eff}$ extracted from the charge measurements under a sinusoidal force excitation of ZnO nanowire-based VINGs and the deduced $g_{33,eff}$ and $g_{33,eff} \cdot d_{33,eff}$, taken from refs. 13 and 27. These FoM analyses allow the evaluation of the devices from different research teams, showing a comparable device performance. In contrast, it is complicated to compare the voltage output signal of GaN nanowire-based VINGs from ref. 51 and 52, under tapping and step-like excitation. This is due to the missing information regarding the impedance characteristics of the device and the detection circuit, which strongly influences the absolute value of the measured voltage signals (see section 3). Importantly, this limited information impedes our access to an insight mechanism responsible for such high voltage output from ref. 52, hindering a possibility to reproduce such encouraging results. It would be helpful to advance the research field of piezo-energy harvesters by providing more details of the devices and the measurement setup as outlined in refs. 22 and 23 as well as in this work. The complete investigations could yield the results that are inherently comparable, offering meaningful information for research communities.

Conclusions and Remarks

This work reports a comprehensive analysis of the piezoelectric response, including the piezo-charge and piezo-voltage of the AlN nanowire-based VINGs as a function of mechanical excitation frequency. We confirm the piezo-origin of the output signal by analysing the phase response of the charge signal in respect to the mechanical excitation, which is consistent with the material polarity of the nanowires measured by other techniques. Piezo-charge, piezo-voltage, and impedance of the devices are directly correlated using an equivalent circuit model. We experimentally demonstrate that comparing solely the absolute values of piezo-voltages from different prototypes can lead to an erroneous interpretation of device performance since such values can vary with testing configurations. We also show that the obtained piezo-voltage is also influenced by the device impedance, which can depend on a variety of parameters. Therefore, it is more reliable to extract the $g$ piezoelectric voltage coefficient and the product $g \cdot d$, both representing important FoMs of piezoelectric materials, to evaluate and compare the device's performance. The piezoelectric $g$ coefficient determined from our results reveal that the nanowire-based VINGs can potentially outperform their rigid counterparts for mechanical sensing and energy generating applications. In addition, without directly measuring the piezoelectric charges generated by the devices, we present alternative possibilities to deduce the $d_{33,eff}$ using the piezoelectric volage and the impedances of the device, along with that of a detection circuit. This work provides an experimental guideline not only for understanding the characteristics of VINGs, but also for facilitating a quantitative comparison between nanostructured piezoelectric sensors and energy harvestors across different architectures within the same and various research groups.

Notes
The authors have no conflicts of interest.

Supporting Information
Equivalent circuits of the piezo-material connected to the detection system, Charge and voltage responses from x-quartz substrate, Dielectric permittivity of a composite layer, High value of piezo-charge response of nanowire-based VINGs, Piezo-charge and piezo-voltage response of AlN nanowire-based VINGs under as the step-like force excitation.

Acknowledgment
We acknowledge the assistance from Nanofab/Néel Institute for sample fabrications and from SERAS/Néel Institute for the set-up development. The work is financially supported by ANR-15-IDEX-02, ANR-PRCI NanoFlex (Project No. ANR-21-CE09-